\documentclass[aps,prl,twocolumn]{revtex4-1}
\usepackage{graphicx}

\usepackage{dcolumn}

\begin{document}


\title{Linear and nonlinear Anderson localization in a curved potential}
\author{Claudio Conti$^{1}$}
\affiliation{
$^1$Department of Physics, University Sapienza, Piazzale Aldo Moro 5, 00185, Rome (IT)}
\email{claudio.conti@uniroma1.it}
\date{\today}

\begin{abstract}
We investigate disorder induced localization in the presence of nonlinearity and curvature.
We numerically analyze the time-resolved three-dimensional expansion of a wave-packet in a bended cigar shaped potential with a focusing Kerr-like interaction term and Gaussian disorder.
We report on a self-consistent analytical theory in which randomness, nonlinearity and geometry are determined by a single scaling parameter,
and show that curvature enhances localization.
\end{abstract}


\maketitle
Nonlinearity, disorder and geometry contribute to various forms of wave-localization. 
In continuous systems, many recent experiments have been reported on Anderson localization, \cite{Swartz07, AspectNature2008, InguscioNature2008, Lodahl2010, Modugno2010, Kondov2011,Lagendijk2012, Maret2013} and several theoretical models have been developed about the role of nonlinearity in disorder-induced states \cite{Kivshar90, Kivshar01b,sanchez-palencia2007, Paul2007, Skipetrov2008, Conti2011, Fishman2012}, even beyond perturbation theory \cite{Conti2012} and including nonlocality \cite{Folli12}.
On the other hand, various authors investigated nonlinear waves in complex geometries, as solitons \cite{batzpeschel2008,batzpeschel2010,peschel2010} and shocks \cite{Conti2013}. 
However, the effect of curvature on Anderson localization and the role of nonlinearity in disordered three-dimensional (3D) manifolds have not been previously considered. 
This topic is expected to be relevant in wave-dynamics in extreme conditions; indeed, nowadays, novel beam shaping and fabrication techniques make
regimes encompassing curved geometry, randomness, and interaction experimentally accessible, as, e.g., in random lasers \cite{leonetti2011,Ghofraniha2013}, quantum fluids \cite{Carusotto2013}, or twisted and bended photonic crystal fibers and optical waveguides \cite{Wong2012,Butsch2012,faccio2012}. In addition, wave-localization in a disordered geometry is also relevant in modern research about quantum-gravity \cite{rothstein2013,quantumgeometrybook}.

In the interplay of disorder, geometry, and nonlinearity, the role of the involved spatial scales is a key issue. 
Without disorder and nonlinearity, strong localization occurs when the radius of curvature is so small to be comparable with the wavelength \cite{daCosta1981,Conti2013}. 
With structural randomness, it may be argued that curvature
is relevant also for shallow topological deformations, because the leading scale is the localization length that may be much larger than the wavelength in waveguides and surfaces embedded in a 3D space, i.e., in one (1D) or two (2D) effective dimensions.

In this Letter, we investigate Anderson localization in a curved potential by the nonlinear Schroedinger (NLS), or Gross-Pitaevskii (GP), equation. 
We show that a deformed geometry enhances the degree of localization due to randomness.
Specifically, we consider a cigar shaped potential in the GP equation, which is bended in the presence of a Gaussian random potential. 
We numerically investigate the 3D+1 spreading of a wave-packet, and trapping due to disorder and nonlinearity. 
The problem is theoretically analyzed, beyond perturbation theory, by resorting to a 1D disorder averaged variational formulation. 
This analysis is valid for an ordered trap superimposed to some disorder, or for a randomly varying shape.
Geometry, nonlinearity, and randomness, are included in a self-consistent way, and results concerning the wave shape, the length of localization and the nonlinear eigenvalue  are derived in closed form. 

\noindent {\it Model ---} In dimensionless units, GP equation reads as
\begin{equation}
i \Psi_t=-\nabla^2 \Psi+V_{3D}(\mathbf{r})\Psi-\chi |\Psi|^2 \Psi\text{,}
\label{GP3D}
\end{equation}
with $N\equiv\int |\Psi^2| d\mathbf{r}$ and $\chi=\pm 1$. We consider a random curved potential, $V_{3D}=V_1(q_1)+V_\perp(q_2,q_3)+V_{R3D}(q_1,q_2,q_3)$ where $q_1=q$ is the {\it longitudinal} coordinate, and $q_{2,3}$  are the {\it transverse} coordinates.\cite{daCosta1981} 
$V_\perp$ determines the transverse confinement, $V_1(q_1)$ is a weak longitudinal trap, 
$V_{R3D}(q_1,q_2,q_3)$ is a Gaussian random potential with $<V_{R3D}(\mathbf{r}) V_{R3D}(\mathbf{r}')>=2D \delta(\mathbf{r}-\mathbf{r}')$ and $D$ measures the strength of disorder.
We specifically investigate an elongated quadratic potential $V_\perp=w (q_2^2+q_3^2)$, $V_1(q_1)=w_1 q_1^2$, with $w>>w_1$, corresponding to a cigar shaped trap in Bose-Einstein condensation (BEC).
The bending is modeled by a parabolic profile of the longitudinal axis of the trap $y=k x^2$; local curvature is given by 
$K(q)=4k^2/\left[1+4 k^2 x(q)^2\right]^{3/2}$,
with $x(q)$ the inverse of $4kq=2kx \sqrt{1+4k^2 x^2}+\sinh^{-1}(2kx)$;
$R_{min}=1/(2k)$ is the minimal radius of curvature at $q=0$ \cite{Conti2013}.

\noindent With reference to BEC, choosing a spatial scale $R_{min}=1$ corresponding to a radius of curvature $R_{MKS}=10\mu$m, and taking an interaction length $a_{MKS}=5$nm, we have for the number of atoms $N_A=10^4 N$, and ground state transversal size of the condensate $l_{\perp,MKS}=4\mu$m.

\noindent {\it Simulations ---}
We numerically solve Eq.(\ref{GP3D}); following the reported experimental investigations \cite{AspectNature2008,InguscioNature2008}, we consider the expansion of a Gaussian wave-packet initially positioned at the bending point $\Psi(\mathbf{r},t=0)=\exp(-r^2/w_0^2)$ with $r^2=q_1^2+q^2_2+q_3^2$ and width $w_0\cong 0.4$, corresponding to the ground state of the transverse trap spreading along the curved potential ($w=100,w_1=10^{-2}$). 

\begin{figure*}
\includegraphics[width=1.0\textwidth]{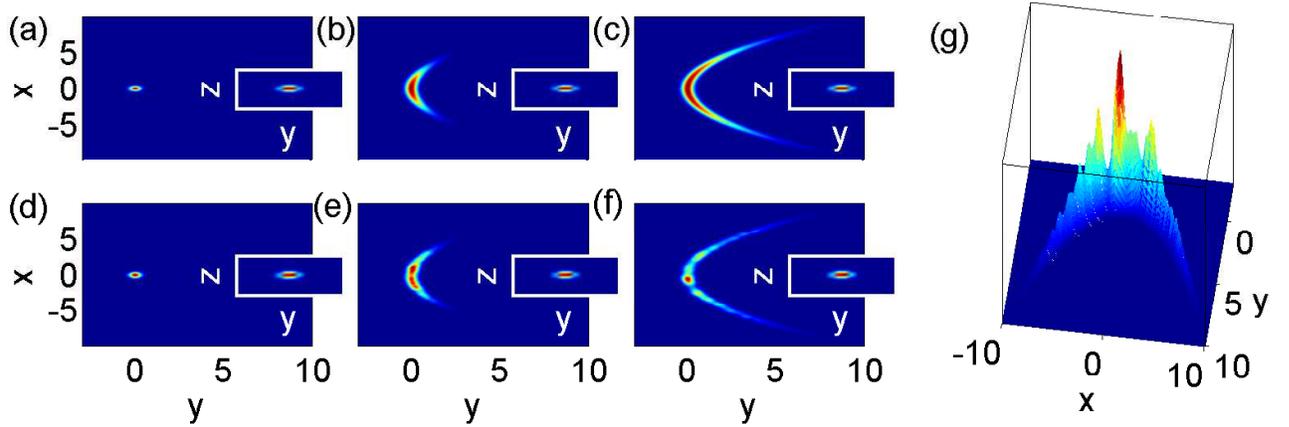}
\caption{(Color online) 
(a,b,c) $|\Psi|^2$ in the absence of disorder ($D=0$, linear case, $\chi=0$) for $t=0,0.5,1$; (d,e,f) as in (a,b,c) with $D=0.05$; (g) 3D surface corresponding to (f). The size of the insets in (a-f) are $\Delta y=\Delta z=4$.
\label{fig1} }
\end{figure*}

We first consider the linear case ($\chi=0$) with low curvature ($k\cong0.1$), and investigate the dynamics without (Fig. \ref{fig1}a,b,c) and with disorder (Fig. \ref{fig1}d,e,f,g). The degree of localization is quantified by the {\it participation ratio} $l_\mathcal{P}$ and its inverse $\mathcal{P}$ 
\begin{equation}
l_{\mathcal{P}}=\frac{1}{\mathcal{P}}=\frac{\left(\int d\mathbf{r} |\Psi|^2\right)^2}{\int d\mathbf{r} |\Psi|^4}.
\end{equation}

\begin{figure}
\includegraphics[width=0.5\textwidth]{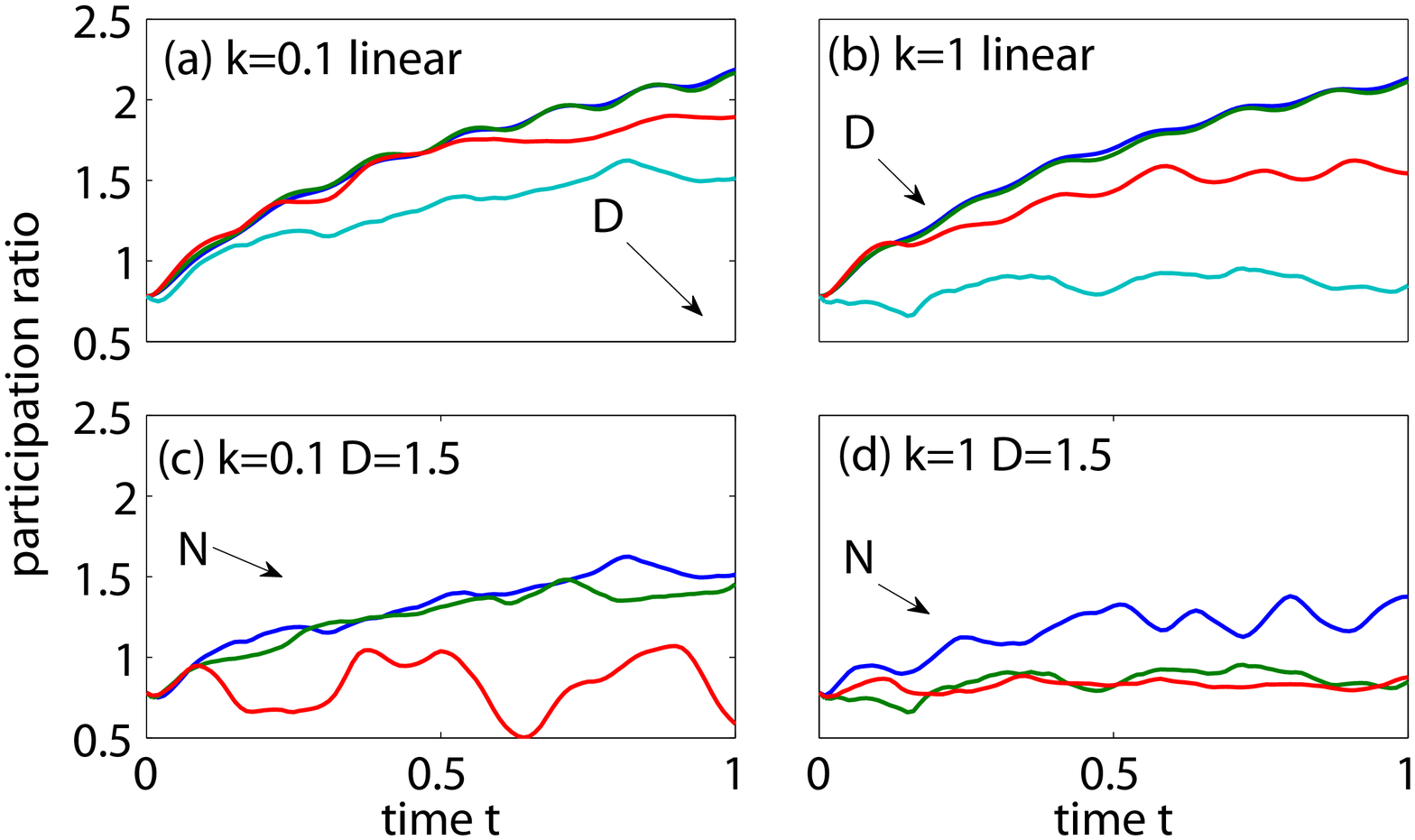}
\caption{(Color online)
Participation ratio $l_\mathcal{P}$ versus time: 
(a) $k=0,1$, $\chi=0$ (linear), and $D=0,0.05,1.0,5$; (b) as in (a) with $k=1$;
(c) $\chi=0$ (linear), $N=0.1$ ($N_A=1000$), and $N=1$ ($N_A=10000$) with $D=1.5$ and $k=0.1$; (d) as in (c) with $k=1$.
\label{fig2} }
\end{figure}

Panels \ref{fig1}a,b,c show that, for $D=0$, the wave spreads along the curved geometry; with disorder (without nonlinearity) the spreading is slowed down by the generation of localizations (\ref{fig1}d,e,f,g). 
For small $D$, $l_\mathcal{P}$  grows with time, but tends to a stationary profile when $D$ increases (Fig.\ref{fig2}a). The localization length is further reduced, and the disorder-induced trapping enhanced, when increasing curvature (Fig.\ref{fig2}b).

\begin{figure}
\includegraphics[width=0.5\textwidth]{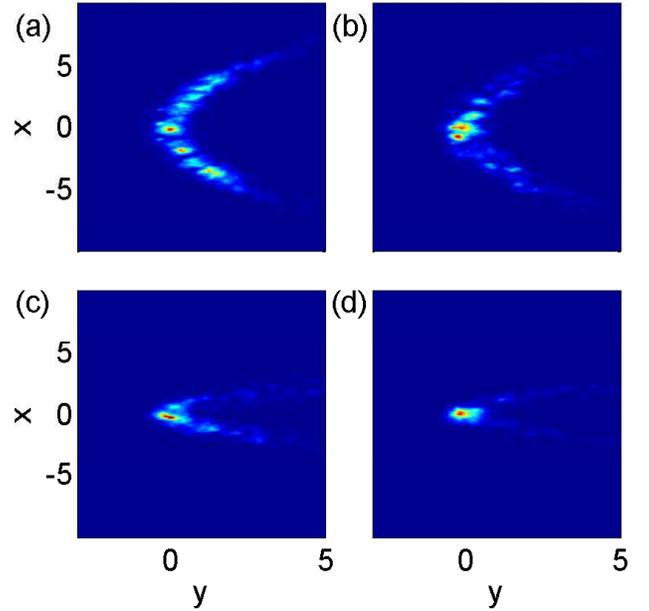}
\caption{(Color online) 
$|\Psi|^2$ at $t=1$ in the bending plane $(x,y)$ for (a) $k=0.1$, $D=1.5$, $\chi=0$ (linear), (b) $k=0.1$, $D=1.5$, $N=0.1$ ($N_A=1000$);
(a) $k=1$, $D=1.5$, $\chi=0$ (linear), (b) $k=1$, $D=1.5$, $N=0.1$ ($N_A=1000$).
\label{fig3}}
\end{figure}

We then consider nonlinearity ($\chi=1$). We show in figure \ref{fig3} 
various profiles $|\Psi|^2$ taken after the expansion has occurred ($t=1$, see Fig.\ref{fig2}), for two curvatures (top and bottom panels), and in the linear and nonlinear cases (left and right panels). In the presence of a focusing nonlinearity $\chi=1$, disorder-induced and nonlinear self-trapping are both enhanced by the bended geometry.
The corresponding participation ratio in Fig.\ref{fig2}c,d shows that the wave displays a quasi-stationary regime, and is more localized when the nonlinearity increases.
In the defocusing case (not reported), we find that, at sufficiently high nonlinearity, any form of localization is destroyed, resulting in curved shock waves \cite{Ghofraniha2012,Conti2013}.

\noindent {\it Theory ---} In the low density regime \cite{Kevrekidis08}, the 3D NLS equation can be reduced to 1D with an effective potential due to curvature \cite{daCosta1981,Conti2013}: when $w/w_1 \rightarrow \infty$, we let
\begin{equation}
\Psi(q_1,q_2,q_3)=l_\perp \psi_\perp(q_2,q_3) \psi(q_1)\exp\left(-i E_\perp t\right)
\end{equation}
with $\int |\psi_\perp|^2(q_2,q_3) dq_2 dq_3= 1$, transverse eigenvalue $E_T$,
transverse localization length $l_T^{-2}=\int |\psi_\perp(q_2,q_3)|^4 dq_2 dq_3$,
and normalization $\int |\psi(q)|^2 dq=N l_\perp^{-2}\equiv P$.
For the parabolic potential, one has for the linear ground state $\psi_\perp=\sqrt{2/\pi l_T^2} \exp(-q_2^2/l_T^2-q_3^2/l_T^2)$ with $l_T=(4/w)^{1/4}$, and $E_T=2 \sqrt{w}$. The resulting 1D NLS equation with geometrical potential $V_G$ is given by
\begin{equation}
i \partial_t \psi=-\partial_{q}^2 \psi+V(q)\psi- \chi |\psi|^2 \psi\text{,}
\label{NLSreduced}
\end{equation}
where $V=V_1+V_G+V_R$, $V_{G}(q)=-K^2(q)/4$,
and $V_R(q)$ a 1D Gaussian random potential such that $<V_R(q) V_R(q')>= V_0^2 \delta(q-q')$  with $V_0^2=2D/l_T^2$.

The 3D+1 numerical results above show that: (i) focusing nonlinearity and (ii) curvature enhance the degree of localization.
Within the 1D reduction, point (i) is explained by known theoretical results 
\cite{Conti2012,Fishman2012}, when neglecting the topology $V_G=0$, while point (ii) can be initially analyzed by standard perturbation theory in the absence of nonlinearity \cite{sanchez-palencia2011} as follows: Eigenstates of (\ref{NLSreduced})  are given by ($V_1=0$)
\begin{equation}
-\varphi_{qq}+[V_R(q)+V_G(q)]\varphi= E\varphi\text{,}
\end{equation}
given a linear localization of $V_R$ with energy $E_n<0$, 
the variation of $E_n$ at the lowest order in curvature is 
\begin{equation}
\Delta E_n=\langle n|V_G|n \rangle=\int \varphi_n^2(q) V_G(q) dq \text{.}
\label{dE}
\end{equation}
As $V_G<0$, the eigenvalue is lowered ($\Delta E_n<0$) by the curved geometry, and the eigenstates become more localized, because, for negative energies, the localization length scales as $l=3/\sqrt{-E}$.  Eq. (\ref{dE}) also implies that localizations located 
at the maximal bending point $q=0$ are more affected by geometry.

Being $\varphi_0=\exp(-2|q|/l_0)/(l_0/2)^{1/2}$ centered at $q=0$, with $l_0$ the average localization length of the linear ground state,
we consider two limits: (a) for a wave-function much more localized than the topological potential, i.e., $l_0<<R_{min}$, we have $\Delta E_n=V_G(0)=-1/(4 R_{min}^2)$ such that the variation of the localization length is $\Delta l/l_0\cong -l_0^2/72 R_{min}^2$ at the lowest order;
(b) when $l_0>>R$, $V_G(q) \cong-\delta(q)/(3 R_{min})$ and 
$\Delta E_0=-\varphi_0(0)^2/(3 R_{min})=-2/(3 l_0 R_{min})$, and $\Delta l/l_0=-l_0/(81 R_{min})$. For both limits the localization length is reduced by an amount growing with $l_0/R_{min}$.

In order to account for nonlinearity, 
we use to the variational approach developed in \cite{Conti2012}. 
Under the so-called annealed approximation the following nonlinear equation for the disorder-averaged bound-state is obtained:
\begin{equation}
-\varphi_{qq}+V_G(q)\varphi-\left(1+\frac{12}{ P l_0}\right) \varphi^3=E\varphi\text{,}
\label{avgsolit1}
\end{equation}
with $E=E(P)$ determined by the condition $\int \varphi^2 dq=P$. Eq.(\ref{avgsolit1}) shows that disorder enhances nonlinearity by a term $\Delta\chi=12/(P l_0)$, and hence favors localization; on the contrary, the larger is the localization length $l_0$ (weaker disorder), the smaller is $\Delta \chi$. In addition, $\Delta \chi$ is negligible  when $P\rightarrow \infty$, corresponding to a dominant nonlinearity with respect to disorder. A related result about the renormalization of disorder by interaction was reported, within a perturbative approach, in \cite{Skipetrov2009}.

\begin{figure}
\includegraphics[width=0.5\textwidth]{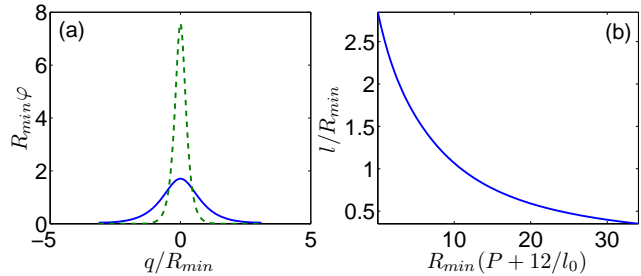}
\caption{(Color online)
(a) Disorder-averaged profile of the nonlinear Anderson localization 
for $R_{min}(P+\frac{12}{l_0})\cong 4$ (continuous line) and $R_{min}(P+\frac{12}{l_0})\cong 21$ (dashed);
(b) nonlinear localization length $l$ versus $R_{min}(P+\frac{12}{l_0})$. 
\label{fig4} }
\end{figure}

To analyze the effect of curvature by Eq.(\ref{avgsolit1}), we consider, as above, the two limits. For (a), $V_G(q)\varphi(q)\cong V_G(0)\varphi(q)=-\varphi(q)/(4 R_{min}^2)$: the effect of the curvature is shifting the nonlinear eigenvalue $E$.
In a flat 1D system \cite{Conti2012}, for the ground state we have  
$E=E_C(P)=-(P^2/16)[1+12/(l_0 P )]^2$ and $l=l_C(P)=3/\sqrt{-E_C}$. With curvature, we have
\begin{equation}
E(P)=E_C(P)-\frac{1}{4 R_{min}^2}\text{,}
\label{energy1R}
\end{equation}
and 
\begin{equation}
l(P)=l_c(P)\left[1-\frac{l_c(P)^2}{72 R_{min}^2}\right]\text{.}
\label{l1}
\end{equation}
Eq.(\ref{l1}) gives the localization length at any power for a finite $R_{min}$.
For the case (b), $V_G(q)=-\delta(q)/(3 R_{min})$ in Eq.(\ref{avgsolit1}), and we have
\begin{equation}
  \varphi(q)=\frac{\sqrt{-2 E(P)}}{\cosh\left[\sqrt{-E(P)}(|q|+\delta l)\right]}\text{,}
\label{solutcurv}
\end{equation}
with $\delta l$ determined by $\sqrt{-E}\tanh(\sqrt{-E}\delta l)=1/(6 R_{min} )$, and measuring the lowering of the localization length. Eq.(\ref{solutcurv}) shows that when increasing the curvature there is a transition from a solitary-wave to an exponentially localized state (see Fig.\ref{fig4}a below).
$E(P)$ is
\begin{equation}
E(P)=E_C(P)\left[1+\frac{l_C(P)}{18 R_{min}}\right]^2 \text{,}
\label{ep1}
\end{equation}
and the energy is reduced ($E_c<0$) by an amount which grows with $l_C(P)/R_{min}$.
The general expression for $l$ is cumbersome and will not be reported, 
at the lowest order in $1/R_{min}$ we have
\begin{equation}
l(P)=l_c(P)\left[1-\frac{5 l_c(P)}{36 R_{min}}\right]\text{.}
\label{dl1}
\end{equation}
Eqs. (\ref{solutcurv},\ref{ep1},\ref{dl1}) give the bound state including all the effects, namely curvature, nonlinearity and disorder.
They imply that the geometrically induced reduction of the localization length is given by the ratio between the localization length and the radius of curvature : the smaller $R_{min}$, the higher the effect on localized states,
which is, however, smaller for highly localized states and for those positioned far from the bending point. 

At variance with the ordered case, where relevant localization is attained for a radius of curvature comparable with the wavelength \cite{daCosta1981,Conti2013}, 
with randomness it is sufficient that the radius of curvature is comparable with the localization length to affect the spatial extension of the states. As the localization length in 1D and 2D may be much larger than the wavelength, topological effects in the presence of disorder are enhanced.

Beyond limits (a) and (b) above, the disorder-averaged bound-state can be found numerically by the solution of Eq.(\ref{avgsolit1}), as in Fig.\ref{fig4}a. 
Scaling arguments in Eq.(\ref{avgsolit1}) show that the eigenvalue $E(P)R_{min}^2$ and localization length $l(P)/R_{min}$ can be written as functions of $R_{min}(P+12/l_0)$,
as reported in figure \ref{fig4}b.
Disorder (measured by $1/l_0$), 
nonlinearity ($P$) and curvature ($R_{min}$) enter as a single parameter $R_{min}(P+12/ l_0)$, which simultaneously accounts for their effects on localization. 
These arguments also applies to the defocusing case.

{\it Conclusions ---}
We investigated the interplay between geometrically induced wave-localization
and nonlinearity for the GP equation with disorder. 
It is found that curvature enhances localization, and all the trapping effects, namely 
disorder, nonlinearity and geometry are determined by a single scaling parameter. Theoretical results
are written in closed form within the annealed phase-space approach,
beyond perturbation theory.
The analysis can be extended to bended 2D surfaces, discrete systems, and other kinds of nonlinearity, as will be reported elsewhere, and show that
curved and twisted structures may  be used to control disorder-mediated effects, as, e.g., random lasing and Anderson localization.

We acknowledge support from the ISCRA High Performance Computing initiative.
The research leading to these results has received funding from the European
Research Council under the European Community's 
Seventh Framework Program (FP7/2007-2013)/ERC grant agreement n.201766.

%

\end{document}